\documentclass[aps,prl,preprint,nofootinbib,superscriptaddress,groupedaddress]{revtex4}  
\usepackage[normalem]{ulem}
\usepackage{amsmath}
\usepackage{enumerate}
\usepackage{amsfonts}
\usepackage{epsfig}
\usepackage{mathbbol}
\usepackage{amsfonts}
\usepackage{dsfont}
\usepackage{color}
\usepackage{subfigure}

\newcommand{\distance}[1]{{\left|\left|{#1}\right|\right|}}

\def\Tr{\mathop{\rm Tr}}
\def\le{\left}
\def\ri{\right}
\def\p{\partial}

\newcommand{\be}{\begin{equation}}
\newcommand{\ee}{\end{equation}}
\newcommand{\ben}{\begin{enumerate}}
\newcommand{\een}{\end{enumerate}}
\newcommand{\bea}{\begin{eqnarray}}
\newcommand{\eea}{\end{eqnarray}}

\newcommand\ket[1]{\ensuremath{\lvert{#1}\rangle}}
\newcommand\bra[1]{\ensuremath{\langle{#1}\rvert}}
\newcommand\sO{{\ensuremath{{\mathcal O}}}}

\newcommand\sE{{\ensuremath{{\mathcal E}}}}
\newcommand\sS{{\ensuremath{{\mathcal S}}}}
\newcommand\Om{\Omega}
\newcommand\vev[1]{{\ensuremath{\left\langle{#1}\right\rangle}}}
\newcommand\ov{\over}
\newcommand\sig{\sigma}
\newcommand\ep{\epsilon}

\newcommand\ha{{1 \ov 2}}

\begin{document}

\preprint{MIT-CTP/4739}

\title{Subsystem Eigenstate Thermalization Hypothesis}

\author{Anatoly Dymarsky} \affiliation{
Department of Physics and Astronomy, University of Kentucky, Lexington, KY 40506, USA\\ 
Skolkovo Institute of Science and Technology,
Skolkovo Innovation Center, 
Moscow  143026
Russia\\ \vspace{0.1pt}}
\author{Nima Lashkari} 
\author{Hong Liu} \affiliation{Center for Theoretical Physics,
Massachusetts
Institute of Technology,
Cambridge, MA 02139}
\date{\today}

\begin{abstract}
Motivated by the qualitative picture of Canonical Typicality, we propose a refined formulation of the Eigenstate Thermalization Hypothesis (ETH) for chaotic quantum systems. The new formulation, which we refer to as subsystem ETH, is in terms of the reduced density matrix of subsystems. This strong form of ETH  outlines the set of observables defined within the subsystem for which it guarantees eigenstate thermalization. We discuss the limits when the size of the subsystem is small or comparable to its complement.  In the latter case we outline the way to calculate the leading volume-proportional contribution to the von Neumann and Renyi entanglment entropies. Finally, we provide numerical evidence for the proposal in the case of a one-dimensional Ising spin-chain.
\end{abstract}

\maketitle

\section{Introduction and Main Results} \label{sec:num}

During the last two decades there has been significant progress in understanding how quantum statistical physics emerges from the dynamics of an isolated quantum many-body system in a pure state.  An important recent development was the realization that a typical pure state, when restricted to a small subsystem, is well approximated by the microcanonical ensemble~\cite{Gemmer, Goldstein, Popescu}. More explicitly, for a system comprised of a sufficiently small subsystem $A$ and its complement $\bar A$, for any random pure state $\Psi$ from an energy shell $(E, E+ \Delta E)$,
\be \label{ope}
\ket{\Psi} = \sum_a c_a \ket{E_a}, \qquad E_a \in (E, E + \Delta E),
\ee
the corresponding reduced density matrix  $\rho_\Psi^A  \equiv  \Tr_{\bar A} \ket{\Psi} \bra{\Psi}$ is almost microcanonical. Taking the average $\vev{\cdots}_\Psi$ over all states~\eqref{ope} with respect to the Haar measure one finds~\cite{Popescu},
\be 
\label{ct}
\vev{||\rho_\Psi^A - \rho^A_{\rm micro} || }_\Psi \leq  \frac{1}{2}{d_A\, \over  \sqrt{d_{\Delta E}}}\ ,\qquad  \distance{O}={1\over 2}\Tr\sqrt{OO^\dagger}\ .
\ee
Here  $ \rho^A_{\rm micro} = \Tr_{\bar A} \rho_{\rm micro}$ is the reduction of the  microcanonical density matrix $\rho_{\rm micro}$ associated with the same energy shell $(E, E + \Delta E)$,
$d_{\Delta E}$ is the number of energy levels inside it, and $d_A={\rm dim}\, \mathcal H_A$ is the dimension of the Hilbert space of $A$.

Equation \eqref{ct} implies that, when the system is sufficiently large, i.e. $\ln d_{\Delta E}\gg \ln d_A$, {the subsystem of a typical pure state is well approximated by that of the microcanonical ensemble with an exponential precision. } 
We refer to this mechanism as ``Canonical Typicality'' (CT). It is important to note that CT is a purely kinematic statement, and provides no insight into whether or how a non-equilibrium initial state thermalizes \cite{footnote1}.

Heuristically, Canonical Typicality can be understood as a consequence of the entanglement between a sufficiently small subsystem and its complement~\cite{Popescu}. While the full system evolves unitarily, a small subsystem can behave thermally as its complement plays the role of a large bath.

Another important development was the so-called Eigenstate Thermalization Hypothesis (ETH)~\cite{Deutsch,Srednicki,Rigol} which conjectures that a chaotic quantum system in a finitely excited energy eigenstate behaves thermally when probed by few-body operators. More explicitly, for a few-body operator $\sO$, ETH postulates that \cite{Srednicki1999,Review}
\be 
\label{ETH0}
\bra{E_a}\mathcal O\ket{E_b}=f_{\sO}(E)\delta_{ab}  +\Omega^{-1/2} (E) r_{ab}\ ,\quad  E=(E_a+E_b)/2\ , 
\ee
where $\ket {E_a}$ denotes an energy eigenstate, $f_\sO (E)$ is a smooth function of $E$,  $\Omega(E) = e^{S (E)}$
is the density of states of the full system, and the fluctuations $r_{ab}$ are of order one, $r_{ab} \sim O(1)$.  The big $O$ here and in what follows refers to the limit when the size of the full system is taken to infinity.

Canonical typicality applies to all systems independent of the Hamiltonian, as opposed to ETH which only concerns chaotic systems, and does not apply to integrable or many-body localized systems. It is a stronger statement, as ETH implies the emergence of the microcanonical ensemble not only for random $\Psi$, but also for a wider class of states, including the linear combination of a few energy eigenstates.

The fact that ETH applies only to chaotic systems 
can be heuristically understood from the general picture of CT; only for chaotic systems energy eigenstates are  ``random enough'' to be typical. 
This perspective thus motivates us to study the properties of the reduced density matrix of a subsystem in an energy eigenstate,  see \cite{Palma,Lai,Garrison:2015lva} for some earlier works.

Now consider a  chaotic many-body system in an energy eigenstate $\ket{E_a}$ 
 reduced to a spatial subsystem $A$ which is smaller than its complement $\bar A$. We postulate the {{\it subsystem ETH}:
\ben 
 \item[(i)]  The reduced density matrix $\rho_a^{A} = {\rm Tr}_{\bar A}  \ket{E_a} \bra{E_a}$ for region $A$ in state $\ket{E_a}$  is exponentially close to some universal density matrix $\rho^A (E)$, which depends {smoothly on} $E$, 
\be \label{eth1}
||\rho_a^{A} -\rho^A(E = E_a )|| \sim O\le( \Om^{-\ha}  (E_a) \ri)\  \qquad \qquad\qquad \quad 
\ee
\item[(ii)] The
``off-diagonal'' matrices $\rho_{ab}^A = {\rm Tr}_{\bar A} \ket{E_a} \bra{E_b}$  are exponentially small,
\be  \label{eth2}
\left|\left| \rho_{ab}^{A}\right|\right| \sim  O\le( \Om^{-\ha}  (E) \ri) 
\ ,\quad   E_a \neq E_b  ,\quad E = \ha (E_a + E_b) \ 
\ee
\een
The pre-exponential factors in (\ref{eth1},\ref{eth2}) could depend on the size of subsystem $A$. Importantly, these factors should remain bounded for the fixed $A$. 
In the next section, we will give numerical support for the exponential suppression of (i) and (ii) using a spin system. } Recently support for (\ref{eth1},\ref{eth2}) was given in the context of CFTs in \cite{Lashkarietal}.

In the thermodynamic limit, i.e.~with the system size taken to infinity, $V\rightarrow \infty$, {while keeping the size of $A$ and the energy density $E/V$ finite and fixed},  
it can be readily seen from  (i) and (ii) that
\be \label{ppe}
\left|\left|\rho_a^{A} - \rho_{\rm micro}^A \right|\right| \sim O(\Delta E/E) \ .
\ee
An implicit assumption here is that $\rho^A (E)$ is well-defined in the thermodynamic limit, i.e.~it  is a function of $E/V$\footnote{At a technical level \eqref{ppe} requires a weaker condition of finite Lipschitz constant $\kappa$ in the thermodynamic limit, $\left|\left| 
\rho^A(E_1)-\rho^A(E_2)\right|\right| \leq \kappa |E_1-E_2|/V$. 
In \eqref{ppe} we also use that for finitely excited states $E\sim V$.} and the prefactor in (\ref{eth1},\ref{eth2}) remains bounded in the limit $V\rightarrow \infty$.
Note that while the suppressions in~\eqref{eth1}--\eqref{eth2} are exponential in the system size, those in \eqref{ppe} are only power law  suppressed.

Using  $\distance{\rho}=\max_{\mathcal O} \Tr(\sO\rho)/2$, 
where maximum is taken over all Hermitian operators of unit norm $\distance{\sO}=1$, 
we conclude from (i) and (ii)  that the matrix elements of $\rho_a^A$ and $\rho^A (E=E_a)$ are exponentially close,
\be
\label{matrixelements}
(\rho_a^A)_{ij} = (\rho^A)_{ij}+O\le(\Om^{-\ha} \ri) \ ,\qquad (\rho_{ab}^{A})_{ij} = O\le( \Om^{-\ha}\ri) \ .
\ee

The formulation in~\eqref{eth1}--\eqref{eth2} is stronger than the conventional form of ETH. In particular, for systems
with an infinite-dimensional local Hilbert space (e.g.~with harmonic oscillators at each lattice site) or continuum field theories  it guarantees ETH for the particular class of observables, while for other observables ETH may not apply. In particular,  subsystem ETH implies the exponential proximity between expectation values in an eigenstate $\bra{E_a}\sO\ket{E_a}$ and the universal value $f_\sO(E_a)=\Tr(\sO \rho^A(E_a))$ for any observable $\sO$ with the support in $A$.\footnote{In case of the continuous quantum field theory, when the full Hilbert space does not admit a tensor product structure of the Hilbert space of $A$ and of its compliment $\bar A$, operators $\mathcal O$ should be defined in terms of the net of local operator algebras, see e.g. \cite{Haag}.} 
This immediately follows from  the Cauchy-Schwarz inequality,\footnote{
For any physically sensible observable $\sO$ the fluctuations of expectation value $\Tr(\rho_a^A \sO)$, which are given by  $\Tr(\mathcal O^2 \rho_a^A)$, must be finite.} 
\be
\label{ETH1}
\Tr((\rho_a^{A} -\rho^A (E_a ))\mathcal O)\leq 2^{1/2}\sqrt{\Tr\left((\rho_a^{A} +\rho^A (E_a )\right)\mathcal O^2)} \distance{\rho_a^{A} -\rho^A(E_a)}^{1/2}\ .
\ee

Moreover, the subsystem ETH can be applied directly to nonlocal measures which are defined in terms of reduced density matrices, such as entanglement entropy, Renyi entropies, negativity, and so on. See e.g.~\cite{Garrison:2015lva} for a recent discussion. In particular, in case of finite-dimensional models it immediately leads to a natural interpretation of thermal entropy as the volume part of the entanglement entropy of a subsystem (see~\cite{Deutsch2,Deutsch3} for recent discussions).  We should caution that  when ${\rm dim}\,\mathcal H_A$ is infinite, arbitrarily close proximity of density matrices does not automatically  imply equality for nonlocal observables. 
For example, in such cases, higher Renyi entropies for $\rho_a^A$ 
may be different from those of the microcanonical or other thermal ensembles~\cite{Lashkarietal}.

In the case of the spin model, for all matrix elements $(\rho^A)_{ij}$, we find strong evidence that $r_{aa}$  of \eqref{ETH0}  are normally distributed . This is consistent with the heuristic picture of typical $\ket{E_a}$ and $r_{ab}$ being a Gaussian random matrix.

It is tempting to ask whether one could further refine pre-exponential factors in~\eqref{eth1}--\eqref{eth2}, especially when the subsystem $A$ is macroscopic.  
Motivated by the $A$-dependent prefactor in~\eqref{ct}  and the average value of the ``off-diagonal" matrices \eqref{eq23}, it is natural to postulate that the pre-factor in~\eqref{eth1}--\eqref{eth2} 
should also be given by 
\be
\label{scaling}
\distance{\rho_a^A-\rho^A (E=E_a)}\sim O\left(e^{N_A -{S (E)  \ov 2}}\right)\ ,\quad \distance{\rho_{ab}^A}\sim
 O\left(e^{N_A  -{S (E)  \ov 2}}\right)\ 
\ee
where $N_A$ denotes the number of effective degrees of freedom in $A$. For a system of finite dimensional Hilbert space, such as a spin system, $e^{N_A}$ simply corresponds to $d_A={\rm dim}\mathcal H_A$, but for a system with an infinite dimensional Hilbert space at each lattice site or a continuum field theory we may view~\eqref{scaling} as a definition of effective number of degrees of freedom. For a spin system we will give some numerical evidence for \eqref{scaling} in the next section.

In addition to~\eqref{ppe} it is interesting to compare $\rho_a^A$ with the reduced density matrices for other statistical ensembles. Of particular interest are the reduced state on the canonical ensemble for the whole system
 \be
 \label{rhoC}
\rho^A_C={\Tr_{\bar A} e^{-\beta H}\over\Tr\,  e^{-\beta H}},
\ee
and  the  local canonical ensemble for the subsystem $A$,
 \be
 \label{rhoG}
\rho_G^A={e^{-\beta H_A}\over \Tr_A\, e^{-\beta H_A}}.
\ee
Here, the Hamiltonian of the subsystem is the restriction of the Hamiltonian $H_A=\Tr_{\bar A}H$.
In~\eqref{rhoC},  $\beta$ is to be chosen so that the average energy of the total system is $E_a$. In~\eqref{rhoG}, $\beta$ can be interpreted as 
a local temperature of $A$ (see also~\cite{ferraro2012,Kliesch,LT}). There is no canonical choice for $\beta$ in this case. Below, we choose it to be the same as in~\eqref{rhoC}. 
In the thermodynamic limit, $V \to \infty$ with the subsystem $A$ and $E/V$ kept fixed,  the standard saddle point approximation argument provides equality between the canonical and the microcanonical ensembles
leading to 
\be \label{1ppe}
\distance{\rho_{\rm micro}^A- \rho_C^A }=O(V^{-1}) \quad \Rightarrow \quad \distance{\rho_a^{A} -   \rho_C^A }=O(V^{-1})\ ,
\ee
where we have also used~\eqref{ppe}. The reduced states $\rho_C^A$ and $ \rho_G^A$ always remain different at the trace distance  level, including thermodynamic limit~\cite{ferraro2012}. Hence,  
\be 
\rho_a^A \neq \rho_G^A ,  \qquad V \to \infty \ .
\ee

Finally, it is interesting to investigate whether~\eqref{eth1}--\eqref{eth2} remain true in an alternative thermodynamic limit when the size of subsystem $A$ scales proportionally with the full system. 
In this limit both the system volume $V$ and the volume $V_A$  for $A$ go to infinity, but we keep the ratio fixed
\be
\label{p-ratio}
0<p={V_A \over V}<\ha\ .
\ee 
Note that for any fixed ratio $p<1/2$ scaling \eqref{scaling} would imply the validity of ETH~\eqref{eth1}--\eqref{eth2}.
In what follows we discuss a weaker version of this statement, which does not rely on \eqref{scaling}. When $A$ is scaled to infinity, we expect $\rho_a^A$ to have a  semi-classical description. 
We conjecture that in this limit $\rho_a^A$ will be approaching $\rho^A(E_a)$ at the level of individual matrix elements,
\bea
\label{me}
(\rho_a^A)_{ij}=(\rho^A(E_a))_{ij}\ .
\eea
Although individual matrix elements will scale as $d_A^{-1}$ and go to zero, \eqref{me} is meaningful as it is satisfied with a precision controlled by $\Omega^{-1/2}\sim d^{-1/2}\ll d_A^{-1}$ for all $p<1/2$.
Furthermore, to the leading order in $1/V$, $\rho^A(E_a)$ will be diagonal in the eigenbasis $\ket{\mathcal E_i}$ of $H_A$, with the diagonal elements given by\footnote{The following form of $\rho_a^A(E_a)$ was previously observed and theoretical justified in~\cite{Tasaki} in the context of a particular model.} 
\be\label{ehne}
\bra{\mathcal E_i}\rho^A(E_a)\ket{\mathcal E_i} = \bra{\mathcal E_i}\rho^A_{\rm mic} \ket{\mathcal E_i} = {\Om_{\bar A} (E_a-\sE_i) \ov \Om (E_a)}\ ,
\ee
where  $\Om_{\bar A}$  is the density of states of $H_{\bar A}=\Tr_A H$. 
 The expression \eqref{ehne} reflects the quasiclassical expectation that the probability to find the subsystem in a state with energy $\mathcal E_i$ is proportional to the number of such states. 
Also for Hamiltonians with local interactions, 
$H = H_A + H_{\bar A}$ up to boundary terms, and in this limit we expect at the level of individual matrix elements 
\be
\label{jj}
(\rho_C^A)_{ij}=(\rho_G^A)_{ij}\ .
\ee
As a self-consistency check, using the expression of $(\rho_a^A)_{ij}$ following from \eqref{me} and \eqref{ehne}, one can calculate $(\rho_C^A)_{ij}$ using saddle point approximation to find that  it is indeed equal to $(\rho_G^A)_{ij}$. Finally note, that in the limit $V\rightarrow \infty$ with $p$ fixed,  $\rho_{\rm micro}^A \neq \rho_C^A$  
and thus we have at the level of individual matrix elements
\be \label{1dif}
\rho_a^A = \rho_{\rm micro}^A \neq \rho_C^A = \rho_G^A\ .
\ee
The form of the density matrix \eqref{ehne} can be used to evaluate leading contribution to the entanglement von Neumann and Renyi entropies in terms of the density of states $\Omega$. This is discussed in Appendix C.
Curiously the leading volume-proportional behavior of the entanglement entropy of $\rho_a^A$ and $\rho_G^A$ is still the same, confirming previous observation of \cite{Garrison:2015lva}, while higher Renyi entropies are different.

In the second part of the paper we provide numerical supports for~\eqref{eth1},~\eqref{eth2}, as well as \eqref{ETH0} and~\eqref{ehne} in a one-dimensional spin chain model.

\section{Numerical results} \label{sec:num}

Now we examine hypothesis~\eqref{eth1} and~\eqref{eth2} of the subsystem  formulation of ETH 
by numerically simulating an Ising spin chain
with a transverse and longitudinal magnetic field 
\be
\label{1dspinchain}
H=-\sum_{k=1}^{n-1}  \sigma_z^k\otimes \sigma_z^{k+1}+g\sum_{k=1}^{n} \sigma^k_x+h\sum_{k=1}^{n} \sigma^k_z.
\ee
This system is known to be non-integrable unless one of the coupling constants $g$ or $h$ is zero.  We solve the system by exact diagonalization for $g=1.05$ and various values of $h$ ranging from $h=0$ to $h=1$. 
For this model, the range of the energy spectrum is roughly from $-n$ to $n$, where $n$ is the total number of spins. The density of states is well approximated by a binomial function, see Appendix A.
We will focus on the behavior of $|E_a\rangle$ for $E_a$ near the central value $E_a\simeq 0$ of the spectrum, which correspond to highly excited states.

\begin{figure}[t]
\subfigure[]{\includegraphics[width=.48\textwidth]{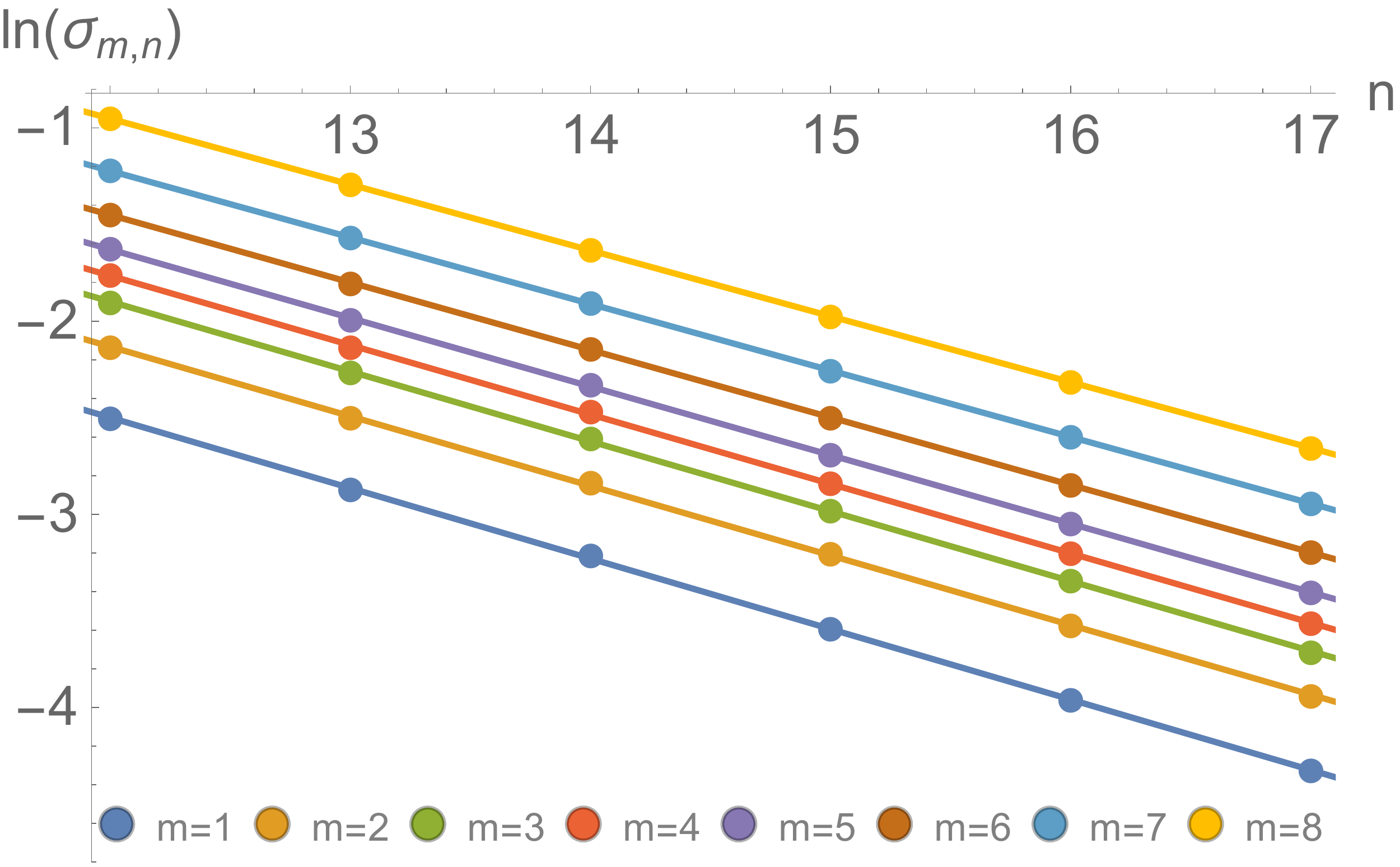}}\hfill
\subfigure[]{\includegraphics[width=.49\textwidth]{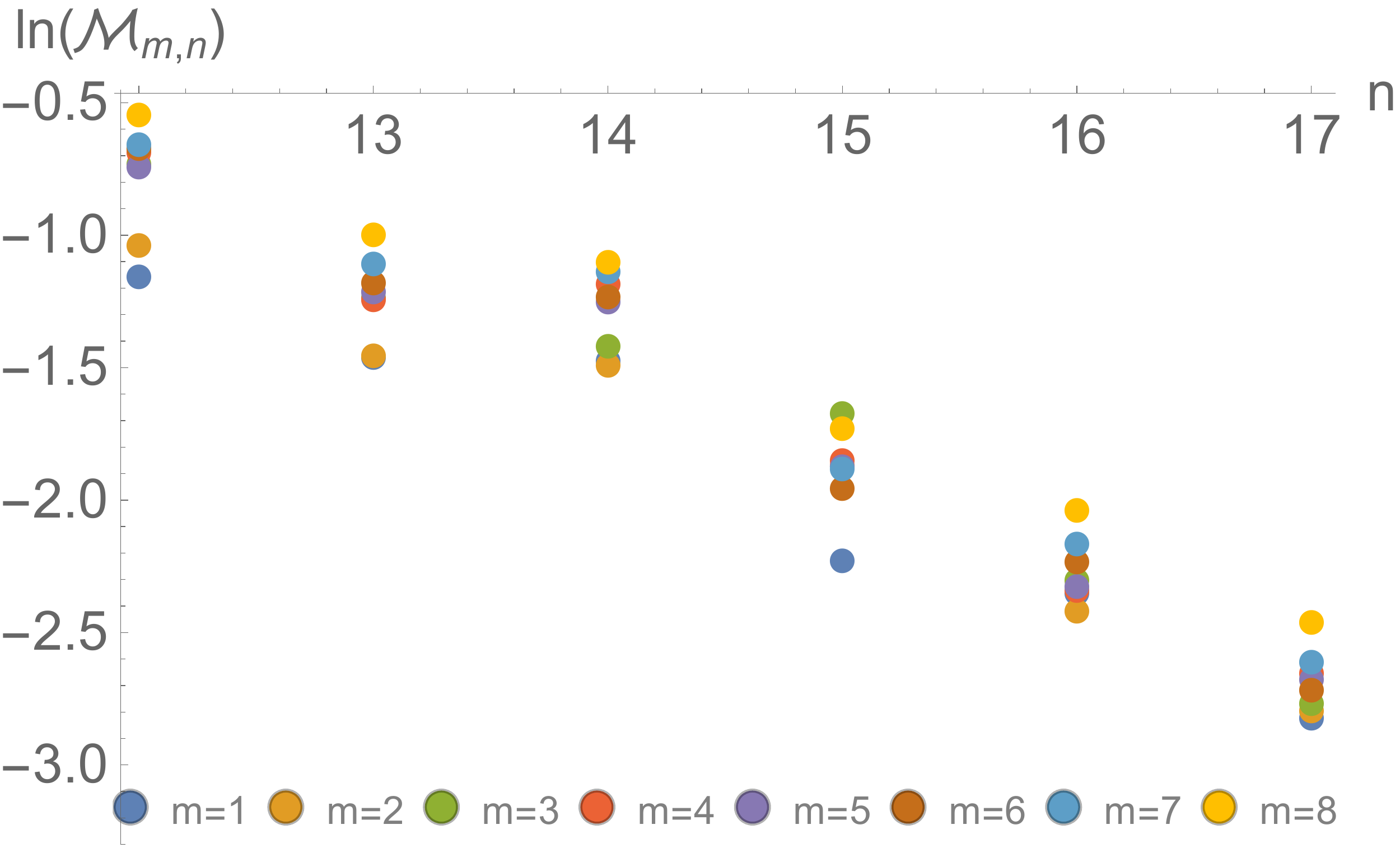}}
\caption{{\bf (a).} Values of $\ln(\sigma_{m,n})$ with superimposed linear fit functions $-\alpha_m n+\beta_m$ for $m=1\dots 8$, $n=12\dots17$ and $\Delta E=0.1n$,  $g=1.05,\ h=0.1$. The slope of linear functions $\alpha_m$ for all $m$ is within $5\%$ close to  the theoretical value $\ln(2)/2$. {\bf (b).} The maximum value of $\Tr(\Delta\rho_a^2)$ over all eigenstates inside the central band $|E_a|\leq \Delta E=0.1n$.}
\label{figure1}
\end{figure}

We denote by $m$ the number of  leftmost  consecutive spins which we take to be subsystem $A$. We introduce the difference between the reduced density matrices for two consecutive  energy eigenstates $\Delta \rho_a=(\rho_{a+1}^A-\rho_{a}^A)/\sqrt{2}$, and define an average variance 
\bea
\label{sigmamn}
\sigma^2_{m,n}={1\over d_{\Delta E}}\sum_a \Tr(\Delta \rho_a^2)\ .
\eea
Here the sum is over all energy eigenstates inside the central band $|E_a|\leq \Delta E$, which is taken to be $\Delta E=0.1n$ and $d_{\Delta E}$ is the total number of states within it. The exponential suppression of $\sigma_{m,n}$ with $n$ is a necessary condition for \eqref{eth1}, as follows from the second inequality below
\bea
\label{distanceinequality}
 \Tr(\Delta\rho^2)\leq 4  \distance{\Delta\rho}^2 \leq d_A \Tr(\Delta\rho^2)\ ,
\eea
valid for any Hermitian $\Delta\rho$ supported on $\mathcal H_A$. Numerical results for $\ln(\sigma_{m,n})$ for different $m$ as a function of total system size $n$ are shown in Fig.\ref{figure1}(a). The  numerical values are well approximated by a linear fit $\ln(\sigma_{m,n})=-\alpha_m n+\beta_m$, with $\beta_m$ increasing with $m$ and the slope $\alpha_m$ for all $m$ being numerically close (within $5\%$ accuracy) to the theoretical value $\ln(2)/2$ suggested by \eqref{eth1}. Similar results are also reported in \cite{Luitz}.

To confirm that \eqref{eth1} for  {\it each} individual $E_a$ is exponentially small, we
examine the maximal value of $\Tr(\Delta \rho_a^2)$ for all $E_a$ within the central band, 
\be 
{\mathcal M}_{m,n}\equiv \max_{a} \Tr(\Delta\rho_a^2)\ .
\ee
The dependence of ${\mathcal M}_{m,n}$ for different $m,n$ is shown in Fig.\ref{figure1}(b). We observe that indeed ${\mathcal M}_{m,n}$ is also exponentially suppressed in $n$.

\begin{figure}[t]
\subfiguretopcaptrue
\subfigure[]{\includegraphics[width=.49\textwidth]{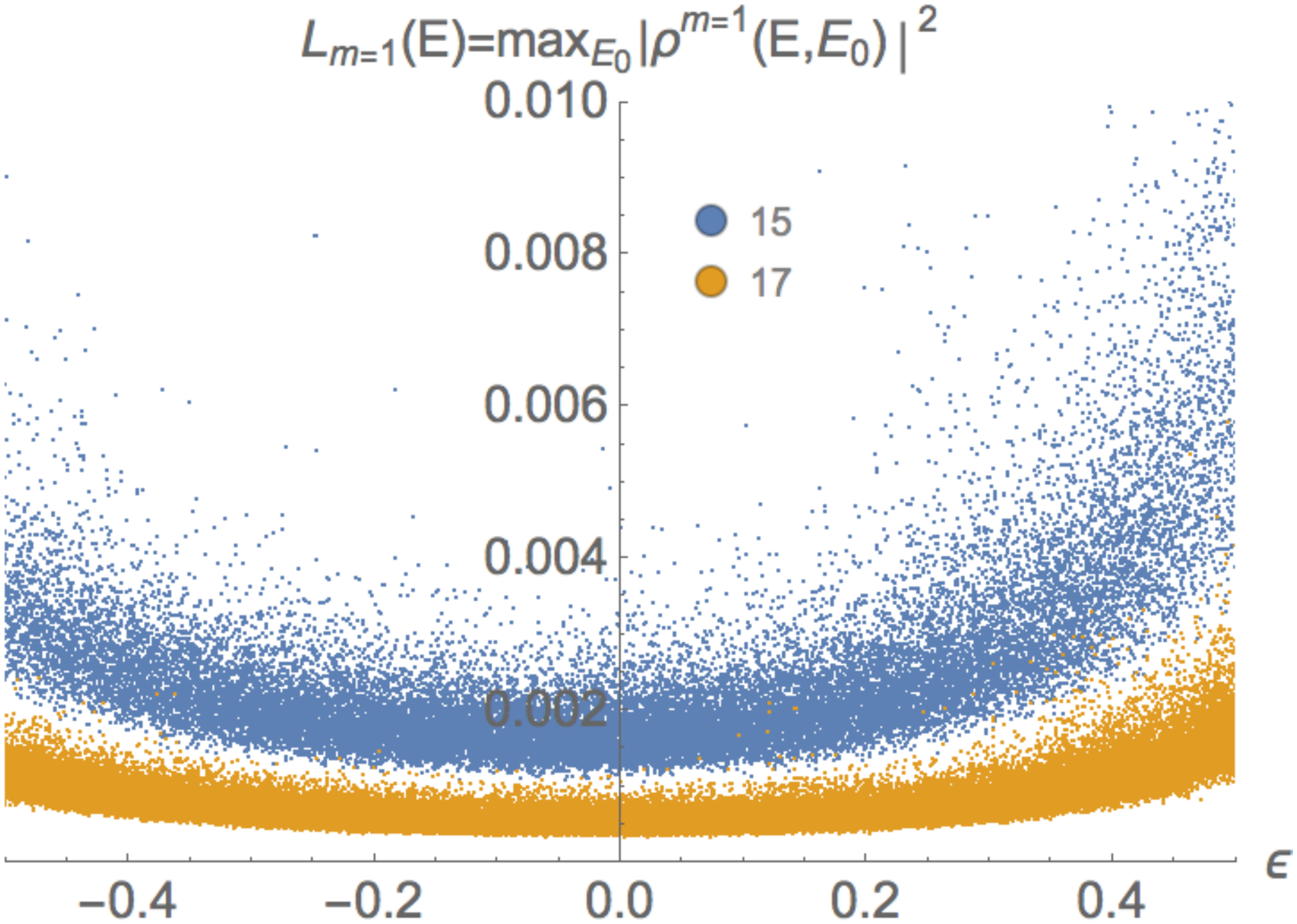}}\hfill \vspace{-10pt}
\subfigure[]{\includegraphics[width=.49\textwidth]{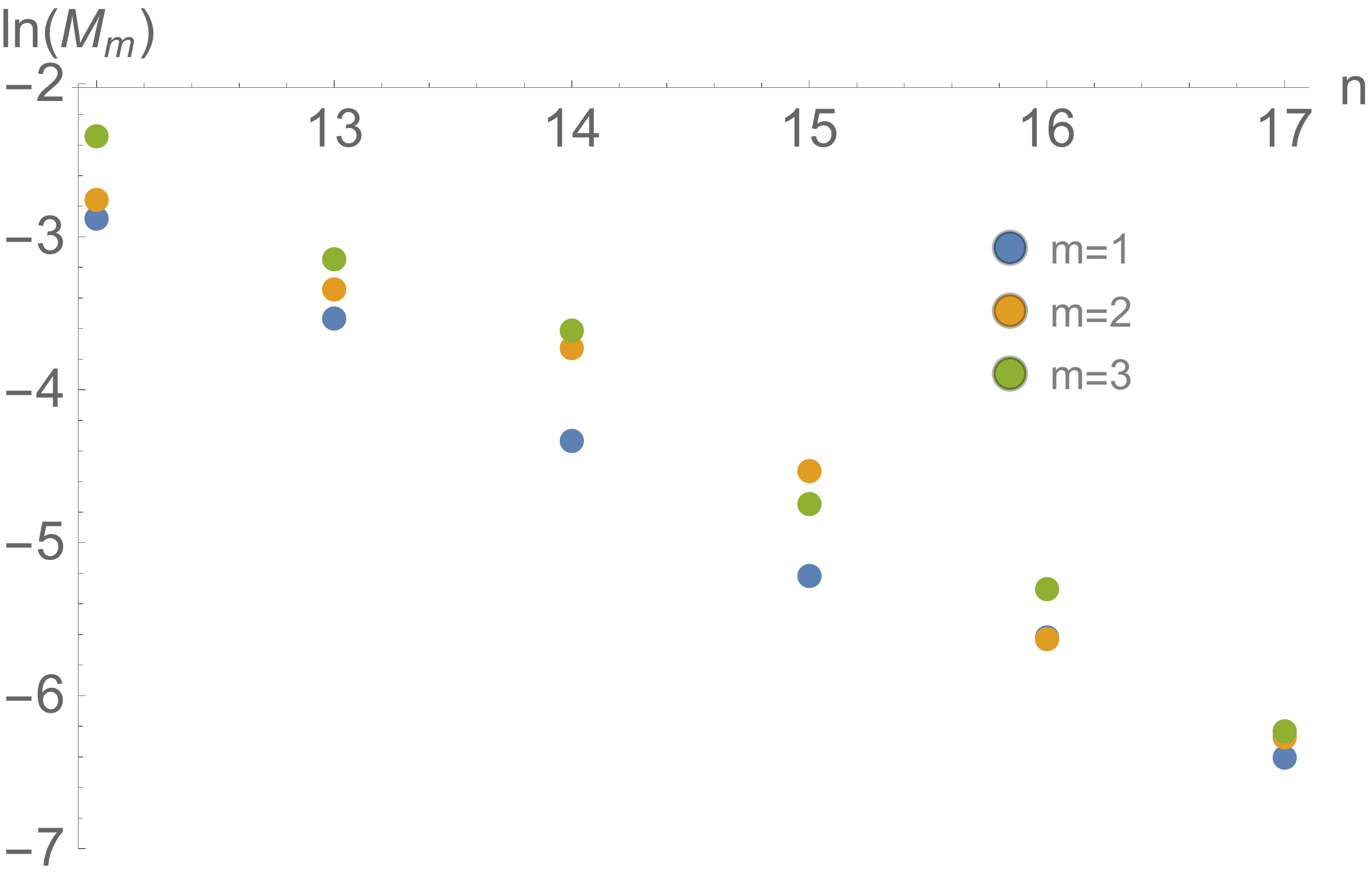}}
\caption{{\bf (a).} Plot of $L_A (E)$ v.s.~$\ep = E/n$ for $n=15$ and $n=17$. {\bf (b).}
$M_{m=1,2,3}$ all decrease exponentially with $n$. Here $\Delta E$ is chosen to be equal $0.1 n$ and $h=0.1$.}
\label{offdiag}
\end{figure}

Now {let} us examine~\eqref{eth2} and show that it is exponentially small for {\it all} sufficiently excited $E_a\neq E_b$. Similar to \eqref{sigmamn}, we consider the mean variance, averaged over all states $E_a$. It can be calculated in full generality for any quantum system (see Appendix B), 
\bea
\label{eq23}
 {1\over d}\sum_{b} \Tr\le((\rho^A_{ab})^\dagger\rho^A_{ab}\ri)={d_A\over d~} \ ,
\eea
where $d$ is the total dimension of the Hilbert space. In the case of spin-chain ${d_A/ d}= e^{- (n-m) \ln 2 }$. This shows that {the} averaged $\distance{\rho_{ab}^A}$ is always exponentially small,  but there remains a possibility 
that a small number of $\Tr(\rho_{ab}^2)$ for $a\neq b$ are actually not suppressed. This is the case for integrable systems. 
To eliminate this possibility, we further examine the following quantity 
\be 
M_{m} \equiv \max_{|E_a|<\Delta E} \max_{b} \Tr((\rho^A_{ab})^\dagger\rho^A_{ab}) ,
\ee
where for a given $E_a$ we first scan all $E_b\neq E_a$ to find the maximal value $L_{ A} (a)\equiv \max_{b} \Tr((\rho^A_{ab})^\dagger\rho^A_{ab})$, and then find  
$M_{A} = \max_a L_{A} (a) $ by scanning all values of $E_a$ within the window $|E_a| < \Delta E=0.1n$.  The restriction to $|E_a| < \Delta E$
is necessary as ETH is only expected to apply to the finitely excited states, not to the states from the edges of the spectrum. This is  manifest from the plot  of Fig.\ref{offdiag}(a). The  plot of   
Fig.\ref{offdiag}(b) indicates that $M_A$ decreases exponentially with $n$. This provides strong numerical support for~\eqref{eth2} and extends the observation of \cite{Kim} that {\it all} diagonal matrix elements satisfy ETH to off-diagonal matrix elements.

\begin{figure}[h]
\subfigure[]{\includegraphics[width=.49\textwidth]{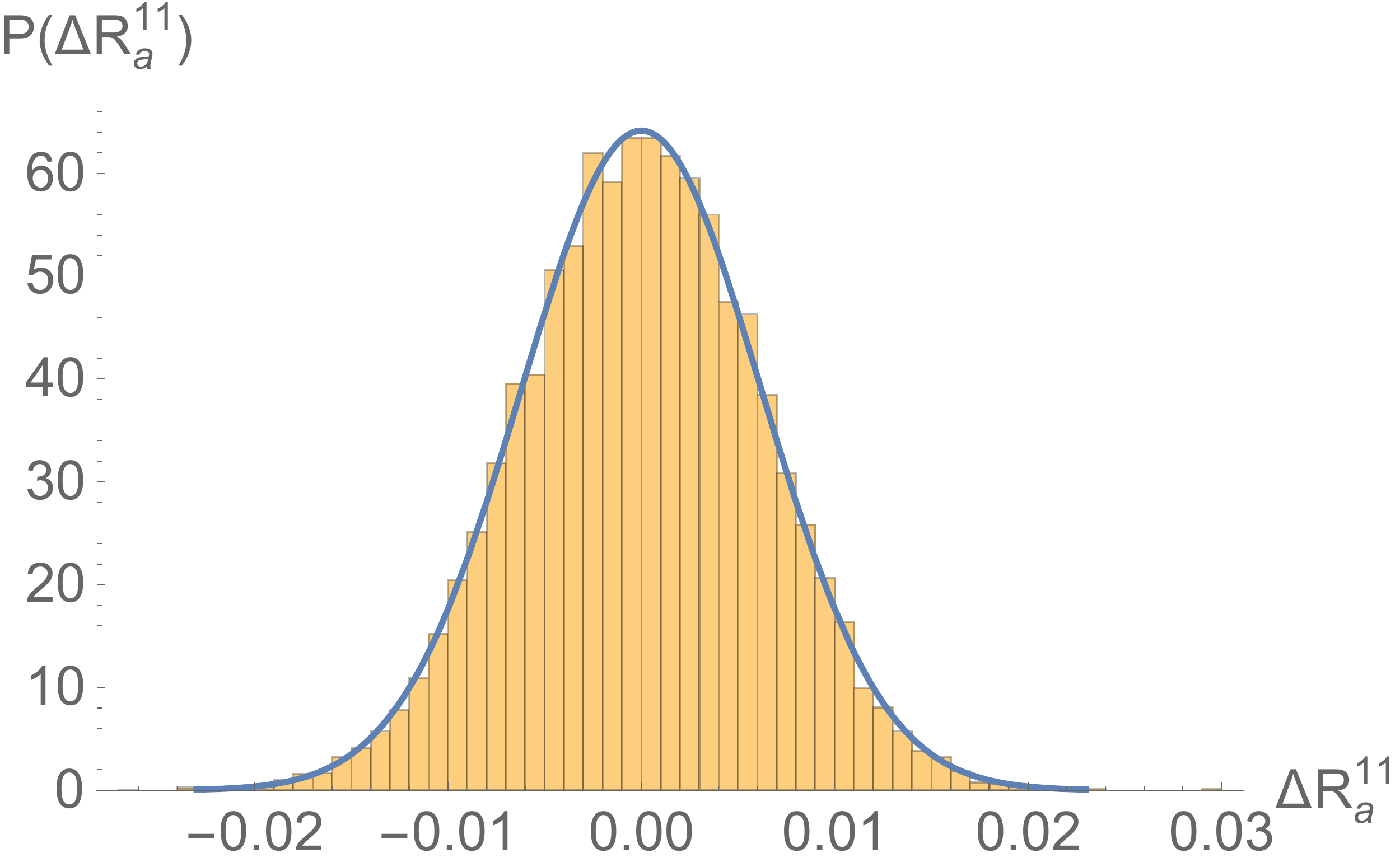}}\hfill 
\subfigure[]{\includegraphics[width=.49\textwidth]{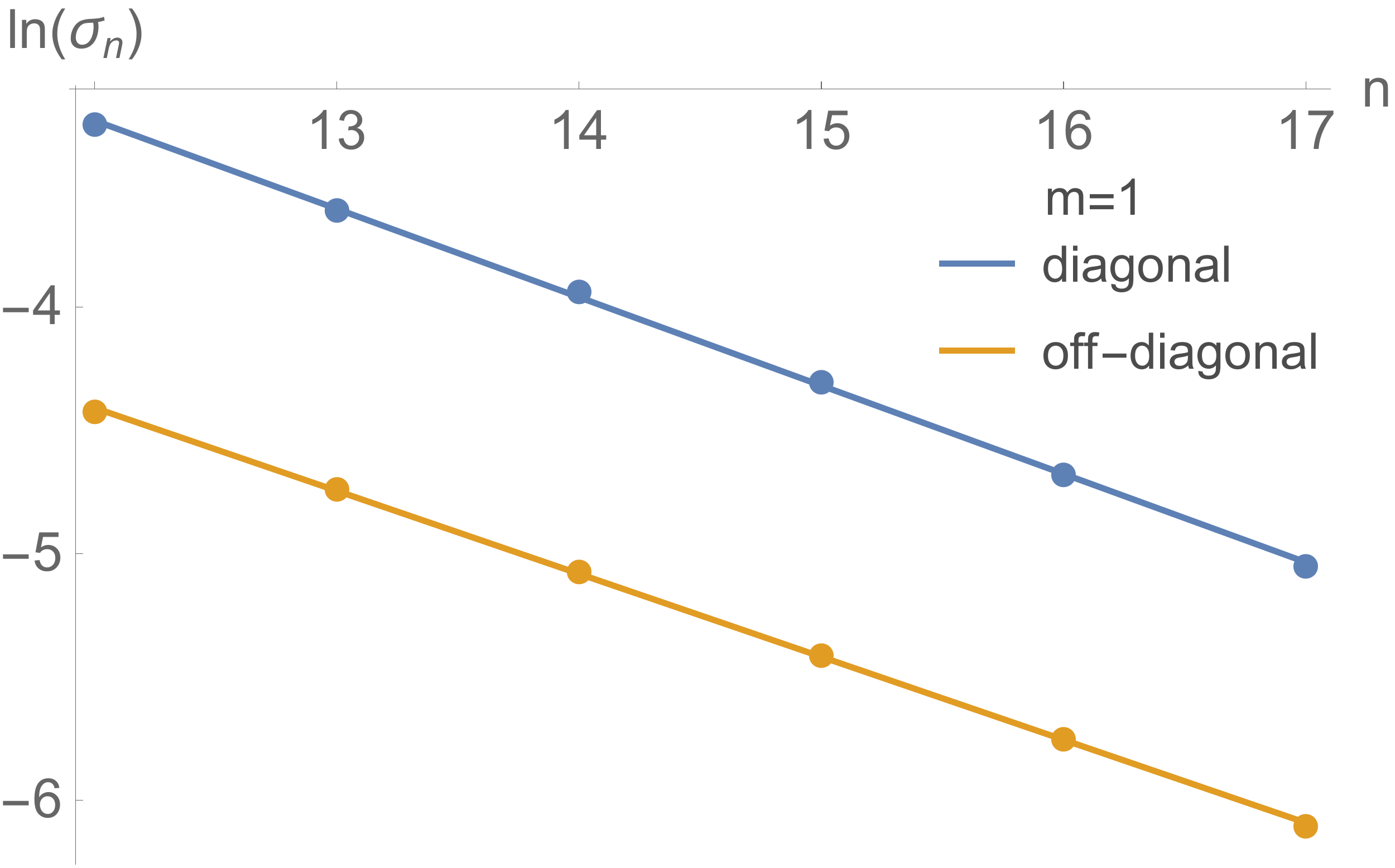}}
\caption{{\bf (a).} Probability distribution $P(\Delta R)$ of the deviation $\Delta R=\Delta R^{11}_a$ corresponding to the matrix element 
$\bra{\mathcal E_1}\rho_a^{m=1}\ket{\mathcal E_1}$ for $\Delta E=0.1n$ and $h=0.1$. It is  superimposed with a Gaussian distribution fit. The vertical axis is the number of energy eigenstates within the energy shell $|E_a|<\Delta E$ with a particular value of $\Delta R$. All matrix elements of $\rho^{m=1,2,3}_E$ show almost identical behavior.
{\bf (b).} Linear behavior of $\ln(\sigma_n)$ as a function of system size $n$ for two matrix elements $\Delta R^{11}$ and  $\Delta R^{12}$  for $m=1$ and $h=0.1$.
Because of the approximate equality $\rho_C\approx \rho_G$ the typical magnitude of the diagonal terms of $\rho_a$ is much larger than the off-diagonal ones. 
There is no qualitative difference between different matrix elements. Results for $m=2,3$ are similar.}
\label{normaldist}
\end{figure}

To study {the} fluctuations $r_{aa}$ of  individual matrix elements of  $\rho^A_E$ around the mean {value} we introduce eigenstates $\mathcal E_{\tilde a}$ of the local Hamiltonian $H_A$ and define 
\be 
\Delta R^{ij}_a={1\over \sqrt{2}}\bra{\mathcal E_i}\rho^A_{a+1} -\rho^A_{a} \ket{\mathcal E_j}\ .
\ee
In terms of the fluctuations $R_{ab}=\Omega^{-1/2}r_{ab}$ of \eqref{ETH0}, $\Delta R_a$ is simply the difference $(R_{(a+1)(a+1)}-R_{aa})/\sqrt{2}$. 
In Fig.\ref{normaldist}(a), we show the distribution (histogram)  $P(\Delta R)$  for $E_a$ from the central band $|E_a|< \Delta E$ and one particular choice of $i,j$ and $A$ consisting of  $m=1$ spin. The plot also contains a superimposed  normal distribution (continuous blue line) that is fitted to have the same variance (and the mean value, which is of order $d_{\Delta E}^{-1}$ i.e.~exponentially small)
\be
\sigma^{ij}_n={1\over d_{\Delta E}}\sum_a (\Delta R_a^{ij})^2\ . 
\ee
Figure\ref{normaldist}(a) shows that $P(\Delta R)$ is well 
approximated by the normal distribution. The situation for all other matrix elements for $m=1,2,3$ is very similar. 

\begin{figure}[t]
\subfiguretopcaptrue
\subfigure[]{\includegraphics[width=.49\textwidth]{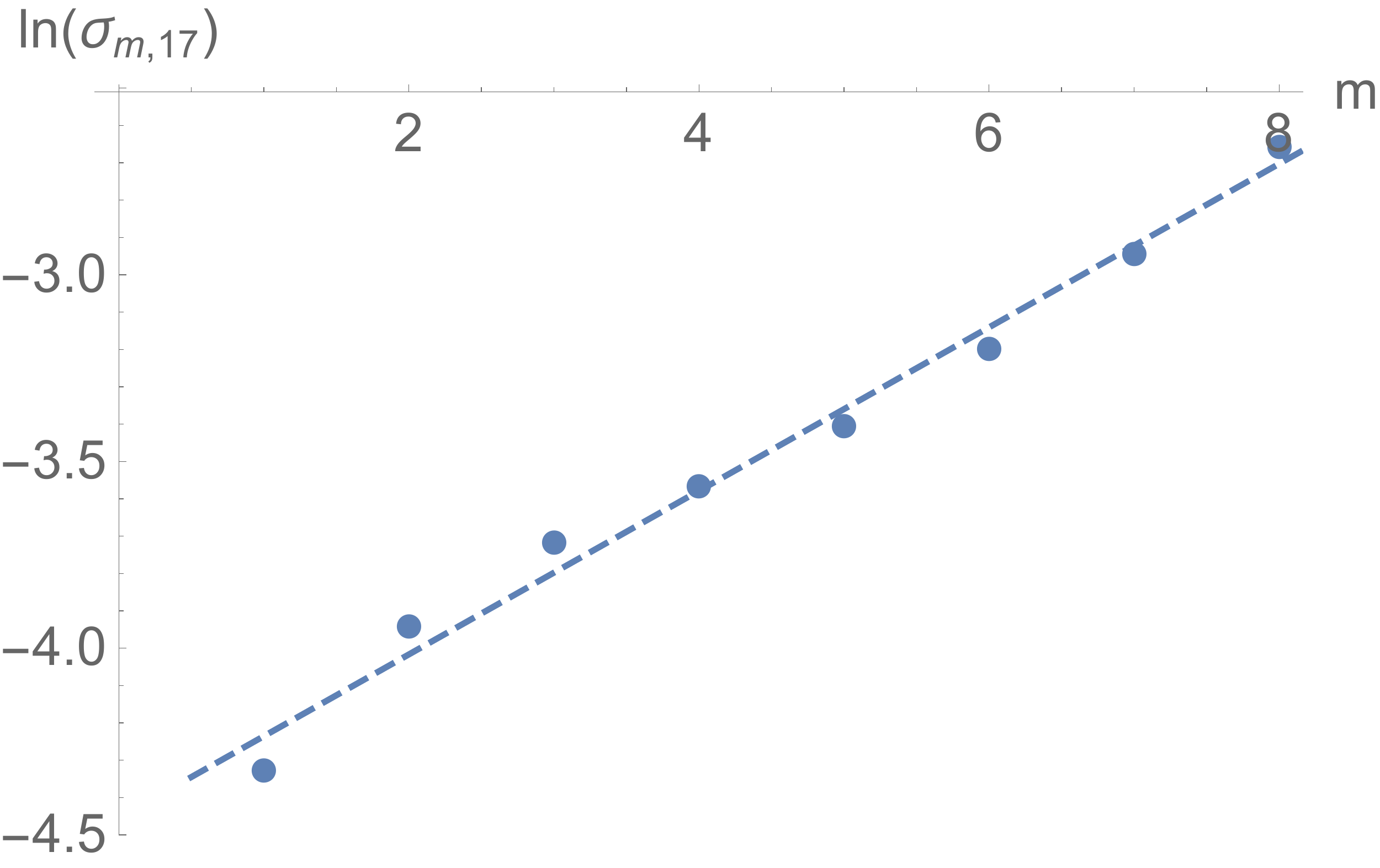}}\hfill
\subfigure[]{\includegraphics[width=.50\textwidth]{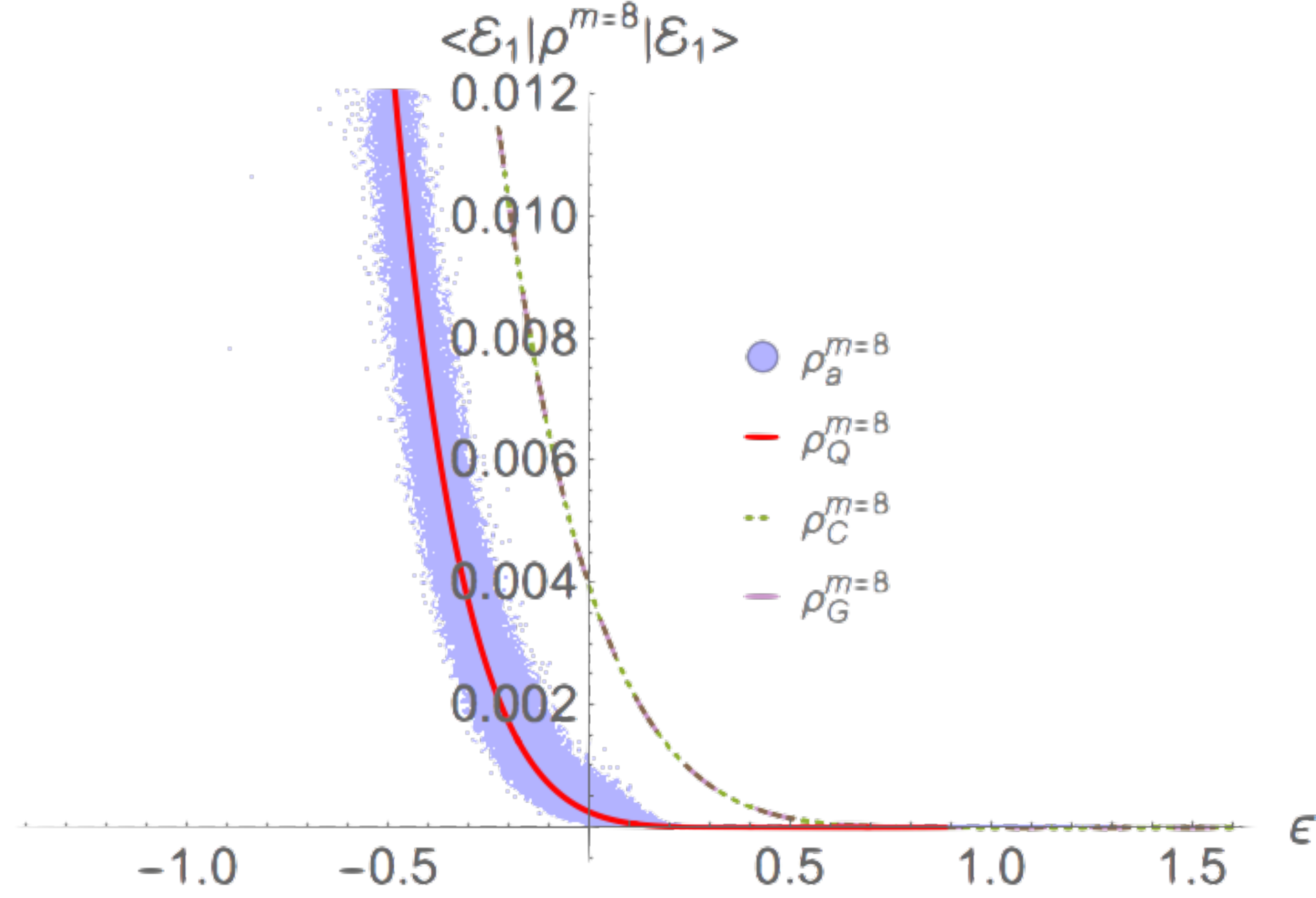}}
\caption{{\bf (a).} Dependence of $\ln\sigma_{m,17}$ on the subsystem size $m$ with the superimposed linear fit $-4.455+ 0.219 m$. {\bf (b).}
Comparison of matrix elements of $\rho_a^A, \rho_C^A, \rho_G^A$ and the quasiclassical result~\eqref{ehne} which we refer to as $\rho_Q^A$.  Blue dots are matrix elements $\bra{\sE_1}\rho^{m=8}_a\ket{\sE_1}$ as a function of energy per site $\epsilon=E_a/n$ for $h=0.1$ and $n=17$. We see that $\bra{\sE_1}\rho_a^{m=8}\ket{\sE_1}$ follows the semi-classical result $\bra{\sE_1}\rho_Q^{m=8}\ket{\sE_1}$ as given by~\eqref{ehne} well, while differs significantly from $\bra{\sE_1}\rho_C^{m=8}\ket{\sE_1} \approx\bra{\sE_1}\rho_G^{m=8}\ket{\sE_1}$,  which lie on top of each other. Quasiclassical result \eqref{ehne} is calculated using density of states $\Omega$ specified in the Appendix A.  
Other matrix elements show similar behavior.} 
\label{quasiclassical}
\end{figure}

Numerically, the standard deviation $\sigma_n$  shows a robust independence of the width of the energy shell $\Delta E$ {that} includes a large  number of states. We plot $\ln(\sigma_n)$ as a function of $n$ in Fig.\ref{normaldist}(b). 
We find that $\sig_n$ decreases exponentially with the system size $n$ for all matrix elements of $\rho^A_a$ for $m =1,2,3$ and values of $h$ which are not too close to the integrable point $h=0$. The exponential suppression of $\Delta R_a$ {follows} from the exponential suppression of $\distance{\Delta \rho}$. But \eqref{eth1}
does not guarantee that different matrix elements of $\rho_a$ would converge to those of $\rho(E_a)$ with the same rate.  Numerics show that the convergence rate $\alpha=d\ln\sigma_n /dn$ is approximately the same, fluctuating around {the} numerical value  $-\ln(2)/2$, for all matrix elements of $\rho^{m=1}_a$, see Fig.\ref{normaldist}(b). The behavior for all matrix elements of $\rho^{m=2,3}_a$ is very similar. {The} numerical proximity of $\alpha$ to  $-\ln(2)/2$ provides a strong numerical support for the form of {the} exponentially suppressed factor in \eqref{ETH0}, which was originally introduced in \cite{Srednicki1999}. 
Provided that $P(\Delta R)$ is well described by  normal distribution,  the probability of a given $R_{aa}$ to be of order $R$ or larger is given by $1-{\rm Erf}(R/\sqrt{2}\sigma_n)\sim e^{- 2^n R^2/R_0^2}$, where $R_0$ is some constant. {If} the total number of eigenstates grows as $2^n$, the probability of finding {an} energy eigenstate $E_a$ which does not satisfy ETH and has large $R_{aa}$  is given by $2^n e^{- 2^n R^2/R_0^2}$. This probability quickly goes to zero with $n$, which explains the strong version of ETH recently discussed in \cite{Kim} and its extension for the off-diagonal matrix elements, which we  observed numerically in Fig.\ref{offdiag}(b).

Next, we investigate the pre-factor in \eqref{eth1} to test the bound behavior outlined in \eqref{scaling}. Namely, we are interested in the dependence of the exponential suppression factor on subsystem size $m$. To illustrate this behavior we plot $\ln\sigma_{m,n}$ for a fixed value of $n=17$ and different $m$ in  Fig.\ref{quasiclassical}(a). In terms of the spin-chain, the bound  \eqref{scaling} means the trace distance $\distance{\rho_a^A-\rho^A(E_a)}$ should not grow faster than $O\left(e^{m \ln(2)-n \ln(2)/2}\right)$. This follows from  \eqref{distanceinequality} if the second norm $\sqrt{\Tr(\rho_a^A-\rho^A(E_a))^2}$ is bounded by $O\left(e^{m \ln(2)/2-n \ln(2)/2}\right)$. The actual slope of the linear fit
of $\ln(\sigma_{m,n})$ as a function of $m$ is $\sim 0.219$. This is substantially smaller than $\ln(2)/2\simeq 0.347$, providing numerical support for \eqref{scaling}.

Finally, we consider the behavior of $\rho^A_a$ when $A$ becomes comparable to $\bar A$ to probe the validity of~\eqref{ehne} in the regime of fixed $p$. 
This is numerically  more challenging. Nevertheless, our numerical results are still quite suggestive. We consider subspace $A$  consisting of $8$ left-most consecutive spins
with $n=17$ and $h=0.1$. The numerical results comparing one diagonal matrix element $\bra{\mathcal E_1}\rho_a^A\ket{\mathcal E_1}$, corresponding to the lowest energy level  of $H_A$ is given in Fig.\ref{quasiclassical}(b). It shows that $\rho_a^A$ follows~\eqref{ehne} pretty well 
while {it} differs from $\rho_C^A \approx \rho_G^A$ significantly.

\vspace{0.2in}   \centerline{\bf{Acknowledgements}} \vspace{0.2in}  This work was supported by  funds provided by MIT-Skoltech Inititative. We would like to thank the University of Kentucky Center for Computational Sciences for computing time on the Lipscomb High Performance Computing Cluster.

\section{\Large \bf Appendix}

\subsection{A. Density of States} 
\label{app:den}
\begin{figure}
\includegraphics[width=.49\textwidth]{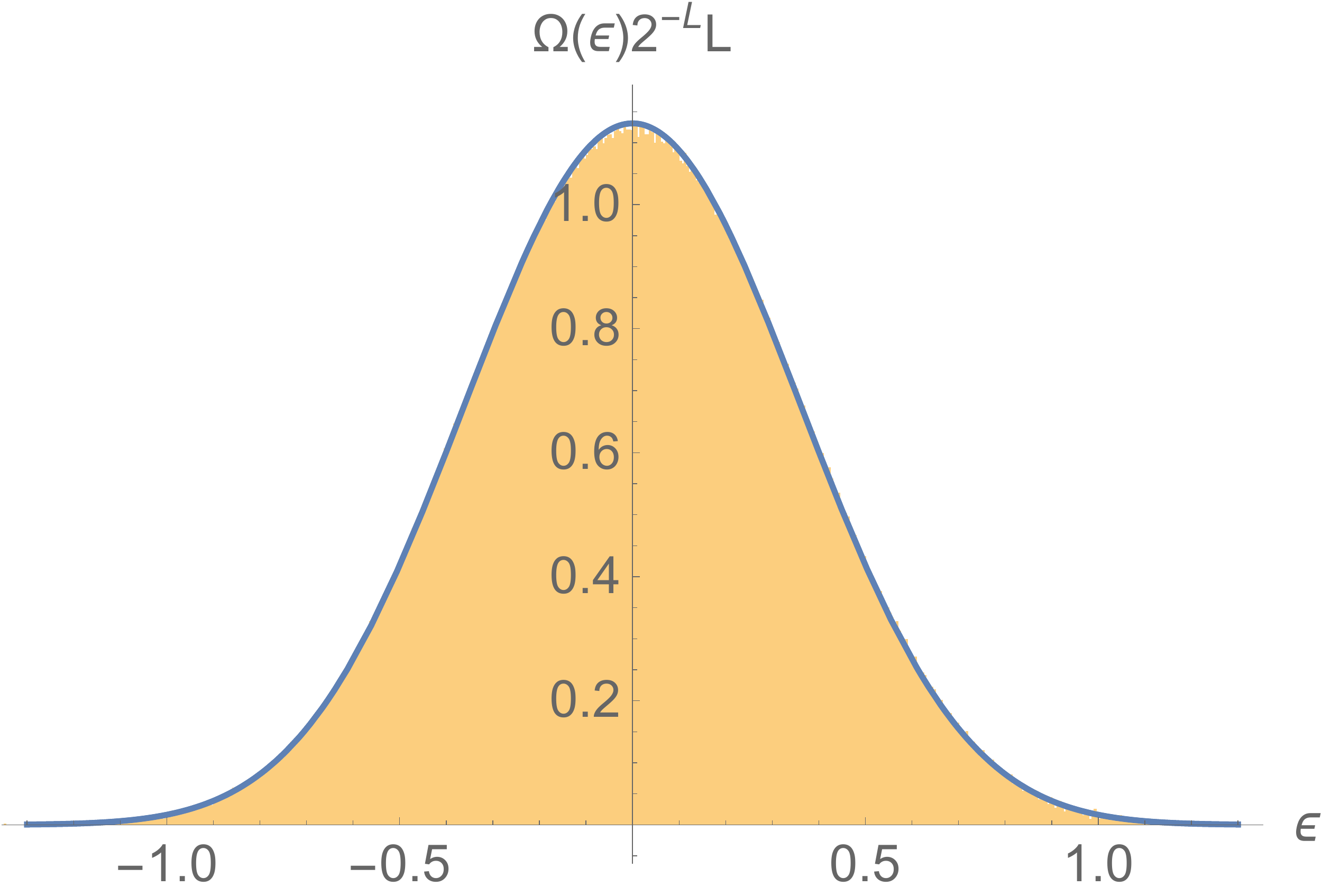}
\caption{The density of states of the spin chain system for $g=1.05, h=0.1, n=17$. The horizontal axis is energy per site $\epsilon=E/n$. 
The yellow bars which fill the plot are the histogram for the density of states calculated using direct diagonalization. The blue solid line is a theoretical 
fit by the binomial distribution function \eqref{omega} with $\kappa\approx 0.3489$, see Appendix A.}
\label{fig:spec}
\end{figure}

A spin-chain without nearest neighbor interactions exhibits a degenerate spectrum with the level spacing of order 1. In this case the density of states is given by the binomial distribution. Once the nearest neighbor interaction term is introduced, the spectrum becomes non-degenerate  with the exponentially small level spacing. In this case the density of states can be described by a smooth function $\Omega(E)$, which would be reasonably approximated by the  binomial distribution. For the spin-chain in question
\be
\label{1dspinchain}
H=-\sum_{i=1}^{n-1}  \sigma_z^i\otimes \sigma_z^{i+1}+g\sum_{i=1}^{n} \sigma^i_x+h\sum_{i=1}^{n} \sigma^i_z\ ,
\ee
we start with the binomial distribution
\bea
\label{omega}
\Omega_n(E)={\kappa\, n!\over (n/2-\kappa\, E)!(n/2+\kappa\, E)!}\ ,
\eea for some $\kappa$, and notice that 
 it is properly normalized for any value of $\kappa$ with an exponential precision, $\int dE\, \Omega_n(E)\simeq 2^n$. We fix the parameter $\kappa$ using the value of the second moment 
\bea
\int dE\, E^2\, \Omega_n(E)\simeq 2^{n-2} {n \kappa^{-2}}=
\Tr H^2.
\eea
The latter could be calculated exactly from~\eqref{1dspinchain} yielding $\kappa={1\over 2} \left(g^2+h^2+1-1/n\right)^{-1/2}$.  The resulting density of states provides a very accurate fit for the exact numerical result as depicted in Fig.~\ref{fig:spec}. The expression for density of states \eqref{omega} is used to calculate $\bra{\mathcal E_i}\rho^A\ket{\mathcal E_i}$ given by \eqref{ehne}, which is shown in Fig.\ref{quasiclassical}(b).

\subsection{B. Variance} \label{app:var}

Consider the variance 
\bea
\Sigma^2_a={1\over d}\sum_{b} \Tr\le((\rho^A_{ab})^\dagger\rho^A_{ab}\ri)
\eea
for some fixed $a$ and $d$ being the dimension of the full Hilbert space. Since $\ket{E_a}$ is a complete basis, 
\be
\label{lemma}
\sum_b \vev{ E_b|\Psi_1} \vev{\Psi_2|E_b}=\vev{\Psi_2|\Psi_1}.
\ee
Now let us introduce a basis in the Hilbert space $\ket{i,\bar j}=\ket{i} \otimes \ket{\bar j}$ associated with the decomposition ${\mathcal H}={\mathcal H}_A\otimes {\mathcal H}_{\bar A}$. Then 
\bea
(\rho^A_{ab})_{ij}=\vev{i|\rho^A_{ab}|j}=\sum_{\bar k} \vev{i,\bar k |E_a} \vev{E_b|j,\bar k}
\eea
and 
\bea \nonumber
\Sigma^2_a={1\over d} \sum_b \sum_{i,j}  \sum_{\bar k, \bar \ell} \vev{i,\bar k |E_a} \vev{ E_b|j,\bar k} \vev{j,\bar \ell |E_b} \vev{ E_a|i,\bar \ell}.
\eea
Now we use \eqref{lemma} to get
\bea
\label{offdiagformula}
\Sigma^2_{a}={d_A \over d~}  \sum_{i}  \sum_{\bar j} \vev{ E_a|i,\bar j}
\vev{ i,\bar j |E_a}={d_A \over d}.
\eea

\subsection{C. Semiclassical expression} \label{app:C}

We now discuss the properties of \eqref{ehne}
\be\label{ehne-appendix}
\bra{\mathcal E_i}\rho_a^A\ket{\mathcal E_i} =\bra{\mathcal E_i}\rho_{\rm micro}^A \ket{\mathcal E_i} = {\Om_{\bar A} (E_a-\sE_i) \ov \Om (E_a)},
\ee
in the limit when 
 \bea
\label{p-ratio-appendix}
0<p={V_A\over V}<\ha
\eea 
 is kept fixed and volume $V\rightarrow \infty$.  
In particular we show that at the leading order in $1/V$ the Von Neumann entropy associated with $\rho_a^A$, which is given by \eqref{ehne-appendix}, is the same 
as for $\rho_G^A$, despite the inequality 
\be \label{1dif}
\rho_a^A = \rho_{\rm micro}^A \neq \rho_C^A = \rho_G^A\ .
\ee

In the limit $V_A\rightarrow \infty$ we can  treat the energy levels $\sE_i$ of $A$ as a continuous variable $\sE$, in terms of which 
\be \label{1o}
\Om (E_a) = \int d \sE \, \Om_A (\sE) \Om_{\bar A} (E_a - \sE) 
\ee
where $\Om_A$ is the density of states for $A$. Now 
introduce 
\be 
\ln \Om_A  \equiv \sS_A , \qquad \ln \Om_{\bar A} \equiv \sS_{\bar A}, \qquad \ln \Om \equiv \sS, 
\ee
with the conventional expectation that the density of states grows exponentially with the volume, 
\be 
\sS_A \propto V_A, \qquad \sS_{\bar A} \propto V_{\bar A}, \qquad \sS \propto V\  .
\ee 

Since both $\sS_A$ and $\sS_{\bar A}$ are proportional to $V$ we can use the saddle point approximation 
in~\eqref{1o} to obtain
\be \label{1c4}
\sS (E) = \sS_A (\bar \sE_A) + \sS_{\bar A} (\bar \sE_{\bar A})
\ee
where $\bar \sE_A$ and $\bar \sE_{\bar A}$ are determined by  
\be \label{1c5}
\bar \sE_A + \bar \sE_{\bar A} = E_a, \qquad {\p \sS_A \ov \p \sE} \biggr|_{\bar \sE_A} = {\p \sS_{\bar A} \ov \p \sE} \biggr|_{\bar \sE_{\bar A}}  \ . 
 \ee
Using saddle point approximation for the canonical ensemble of the whole system we recover the conventional relation between the inverse temperature $\beta$ and the mean energy $E$, 
\be 
\beta = {\p \sS(E_a) \ov \p E}. 
\ee 
Together with~\eqref{1c4}--\eqref{1c5} this implies
\be 
\beta = {\p \sS_A \ov \p \sE} \biggr|_{\bar \sE_A} = {\p \sS_{\bar A} \ov \p \sE} \biggr|_{\bar \sE_{\bar A}} \ .
\ee
Then it follows in a standard way that the entropy $S_G^A$ associated with the diagonal density matrix $\rho_G^A$
\bea
\label{rhog}
\langle \sE|\rho_G^A|\sE\rangle\simeq e^{-\beta(\sE-\bar \sE_A)-\sS_A (\bar \sE_A)}
\eea 
is simply $S_G^A = \sS_A (\bar \sE_A)$. 

With help of~\eqref{1c4} one can rewrite \eqref{ehne}, \eqref{ehne-appendix} as follows,
\be  \label{1rg}
\langle \sE|\rho_a^A  |\sE\rangle  \simeq e^{\sS_{\bar A} (E - \sE) - \sS_{\bar A} (E - \bar \sE_A) - \sS_A (\bar \sE_A)}\ ,
\ee 
while off-diagonal matrix elements are negligible.  Then the corresponding entropy $S_a^A$ is given by
\be 
S_a^A = - \Tr_A \rho_a^A \ln \rho_a^A =\sS_A (\bar \sE_A)= S_G^A \ .
\ee
Applying saddle point approximation to powers of~\eqref{rhog} and \eqref{1rg} one can readily calculate leading volume-proportional contribution to Renyi entropies for $\rho_a^A$ and $\rho_G^A$ and see that they are different. This is consistent with the numerical results of \cite{Garrison:2015lva}.

\bibliographystyle{unsrt}

\end{document}